\documentclass[pre, preprint, showpacs,preprintnumbers, a4paper]{revtex4-1}
\usepackage[utf8x]{inputenc}
\usepackage[english]{babel}
\usepackage{graphicx}
\usepackage{dcolumn}
\usepackage{amsmath}
\usepackage{amssymb}
\usepackage{bm}
\begin{document}
 \title{Time-delayed feedback control of breathing localized structures in a three-component reaction-diffusion system}
\author{Svetlana V. Gurevich}
\email{gurevics@uni-muenster.de}
\affiliation{Institute for Theoretical Physics, University of M\"unster,
Wilhelm-Klemm-Str.\,9, D-48149 M\"unster, Germany\\
}
\date{\today}
 \begin{abstract}
  We investigate the dynamics of a single breathing localized structure in a three-component reaction-diffusion system subjected to the time-delayed feedback. We show that variation of the delay time and the feedback strength can lead either to stabilization of the breathing or to delay-induced periodic or quasiperiodic oscillations of the localized structure. We provide a bifurcation analysis of the system in question and derive an order parameter equation, which describes the dynamics of the localized structure in the vicinity of the Hopf bifurcation. With the aid of this equation, boundaries of the stabilization domains as well as the dependence of the oscillation radius on delay parameters can be explicitly derived, providing a robust mechanism to control the behavior of the breathing localized structure in a straightforward manner.
 \end{abstract}
 
 
 \maketitle
 \section{Introduction}

Starting with the work of Ott, Grebogi and Yorke~\cite{Ott}, a variety of different techniques for controlling unstable or chaotic states in complex systems has been developed within the last decade. Among other control methods, time-delayed feedback scheme~\cite{PyragasPRA1992} (also referred to as Pyragas control or time-delay autosynchronization) proves oneself to be an efficient tool, allowing a noninvasive stabilization of unstable periodic orbits of dynamical systems~(see, e.g.,~\cite{SchoellChaos2009} and references therein). In the meantime, time-delayed feedback control has been successfully applied to a broad variety of spatially extended systems, including, e.g., plasma physics~\cite{Friedel1998}, nonlinear optics~\cite{PaulauPRE2008, TlidiPRL2009, Green2009, Pimenov2013}, electrochemical~\cite{Kehrt2009} and neural systems~\cite{DahlemCahos2008, Dahlem2010}. In particular, control of the dynamics of spatiotemporal patterns in reaction-diffusion systems have been of increasing interest in recent years. We mention  only delay-induced turbulent structures in a diffusive Hutchinson equation~\cite{Bestehorn2004}, delay-modulated oscillatory hexagon superlattices and stripes in a Brusselator model~\cite{Li2007}, control of spatiotemporal patterns in a Gray-Scott model~\cite{Kyrychko2009}, spatiotemporal patterns in a prey-predator plankton system~\cite{TianPRE2013}, dynamics of Turing patterns in the Lengyel-Epstein system under time-delayed feedback~\cite{Iri2014} or moving localized and periodic structures in the FitzHugh-Nagumo model~\cite{TlidiPRE2013}. Quite recently, the influence of the delayed feedback on the stability properties of a single stationary localized structure in a three component reaction-diffusion system was studied in details~\cite{GurevichPRE2013}. It was shown that the presence of the feedback force can induce complex dynamical behavior of the localized solution, leading among other things to the formation of moving or breathing structures. However, an opposite problem of stabilization of a certain unstable localized state is still not understood to a large extent.

In this paper we are interested in the influence of the time delayed control on the dynamics of breathing localized structures in a three-component reaction-diffusion system with one activator and two inhibitors. We shall show that a variation of the delay time and the feedback strength can indeed lead to stabilization of the breathing. In addition, more complex delay-induced periodic or quasiperiodic oscillations of the localized structure can also be found. In order to understand the impact of the delayed feedback term on the dynamics of the breathing localized structure, we derive an order parameter equation in the vicinity of the bifurcation point where oscillatory dynamics sets in. The desired equation is a subject to a nonlinear delay-differential equation, explicitly describing the temporal evolution of the localized structure.

\section{Linear stability analysis}

As mentioned above, here we are interested in the dynamics of breathing localized structures in a three-component reaction-diffusion system with one activator and two inhibitors:
 \begin{equation}
    \begin{split}\label{eq:3krd}
      \partial_{t}u&=D_u\,\Delta u+f(u)-\kappa_3\, v-\kappa_4\, w+\kappa_1+\alpha\,\bigl(u(t)-u(t-\tau)\bigr)\,,\\
      \eta\,\partial_{t}v&=D_v\,\Delta v+u-v+\eta\,\alpha\,\bigl(v(t)-v(t-\tau)\bigr)\,,\\
      \theta\,\partial_{t}w&=D_w\,\Delta w+u-w+\theta\,\alpha\,\bigl(w(t)-w(t-\tau)\bigr)\,.
    \end{split}
    \end{equation}
 Here $u=u(\mathbf{r},t)$ is the activating component, whereas $v=v(\mathbf{r},t)$ and $w=w(\mathbf{r},t)$ denote the inhibiting components, $\mathbf{r}\in\mathbb{R}^2$. The coefficient $\lambda$ in the polynomial nonlinear function $f(u)=\lambda\, u-u^3$ is positive as well as diffusion coefficients $D_u,\,D_v,\,D_w$ of the corresponding components and dimensionless constants $\eta$ and $\theta$, representing the ratios of the characteristic time scales of both inhibitors $v$ and $w$ with respect to that of the activator. The constants $\kappa_3$ and $\kappa_4$ are also positive, whereas $\kappa_1$ violates the inversion symmetry and has arbitrary sign. Finally, $\tau$ denotes the delay time, whereas the parameter $\alpha$ is the delay strength. Note that the time delayed feedback term is introduced in such a way that the corresponding coupling matrix is an unit one~\cite{GurevichPRE2013}. In the absence of the delayed feedback the system~(\ref{eq:3krd}) was first introduced in~\cite{Pu48, Pu53} as an 
extension of the phenomenological model for a planar dc gas-discharge system with high-ohmic semiconductor electrode. On the other hand, the system~(\ref{eq:3krd}) can be considered  in a more general contexts, like as a three-component extension of the FitzHugh-Nagumo system for nerve pulse transmission~\cite{PuBook, Kapral1995, Dahlem2010, PurwinsDS2010} or a model system of a pattern forming chemical reaction in a microemulsion~\cite{Cherkashin2008, AlonsoJCP2011}. However, in the present study we have no specific application in mind and provide a general analysis of the system in question which can be later addressed to a specific applications mentioned above. Notice that from a practical perspective, the usage of identity control scheme may be quite restrictive. Notwithstanding, a close analytic treatment of the simplest case of the control scheme enables to gain a deeper insight into underlying stabilization mechanisms and to get some ideas about the impact of the time delayed feedback on the dynamical 
properties of the localized structure.

From now on we consider the general form of the reaction-diffusion system~(\ref{eq:3krd}) 
\begin{equation}\label{eq:GenEqDel}
  \partial_t\mathbf{q}(\mathbf{r},t)=\mathfrak{L}[\mathbf{q}(\mathbf{r},t)]+\alpha\,\mathrm{E}\,\bigl(\mathbf{q}(\mathbf{r},t)-\mathbf{q}(\mathbf{r},t-\tau)\bigr)\,,
 \end{equation}
 where  $\mathbf{q}=\mathbf{q}(\mathbf{r},t)=(u(\mathbf{r},t)\,,v(\mathbf{r},t)\,,w(\mathbf{r},t))^T$ is a vector-function, $\mathbf{r}\in \mathbb{R}^2$,  $\mathfrak{L}=\mathfrak{L}(\nabla)$ is a nonlinear reaction-diffusion operator and $\mathrm{E}$ denotes an identity coupling matrix. We are interested in the dynamics of a two-dimensional stationary single localized solution $\mathbf{q_0}(\mathbf{r})$ of the system~(\ref{eq:GenEqDel}), which exists and which is stable in appropriate range of parameters~\cite{HeijsterSandstedte2011}. In the absence of the delay term, i.e., for $\alpha=0$, this stationary localized structure can lose its stability with the change of one or more control parameters, e.g., $\eta$ or $\theta$. Typical destabilization scenarios include drift-bifurcation, leading to a motion of localized structures~\cite{OrGuilPRE98,Gurevich2004}, the Hopf-bifurcation (also called Andronov-Hopf bifurcation), where so-called breathing localized  structures are formed~\cite{GurevichPRE2006} or a 
nontrivial combination of both instabilities in the vicinity of a codimensional-two bifurcation point~\cite{GuFrMMNP13}. 
 
\begin{figure}[!h]
\centering
\includegraphics[width=0.6\textwidth]{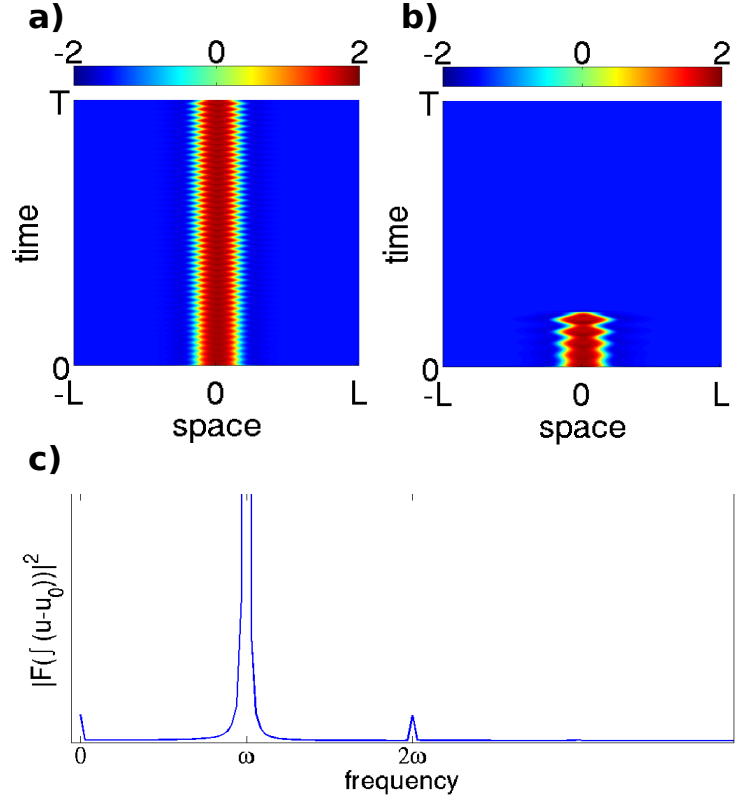}
\caption{Space-time plots for different control parameter values obtained from numerical solution of Eq.~(\ref{eq:3krd}) in the absence of the time-delayed feedback. Time evolution of  the cross-section of the initial pulse close to the stationary solution is shown. (a) $\eta=0.6$, $\theta=0.58$: Amplitude of oscillations reaches a constant value; (b) $\eta=0.62$, $\theta=0.64$: Increase of the control parameter beyond the critical value leads to the collapse of the localized solution. (c) Power spectrum $|F(\int_{\Omega}(u(\mathbf{r},\,t)-u_0(\mathbf{r})))|^2$, calculated for the activator distribution of the breathing localized structure, shown in (b). Here $u_0$ denotes a stationary distribution of the activator field. Other parameters:$D_u=4.7\cdot10^{-3}$, $D_v=0$, $D_w=0.01$, $\lambda=5.67$, $\kappa_1=-1.04$, $\kappa_3=1.0$, $\kappa_4=3.33$, $\alpha=0$, $\tau=0$. The calculations were performed on the rectangular domain  $\Omega=[-L,\, L]\times[-L,\, L]$, $L=1.0$ with periodic boundary conditions.}
\label{fig1}
\end{figure}

\begin{figure}[!h]
\centering\includegraphics[width=0.45\textwidth]{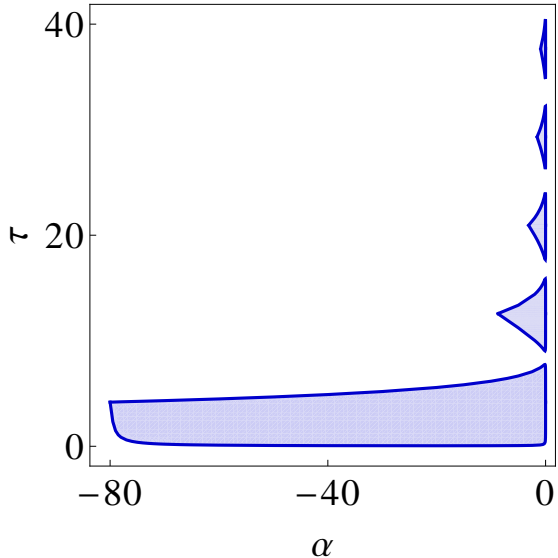}
\caption{Bifurcation diagram in $(\alpha\,,\tau)$ plane obtained from numerical solution of the solvability condition~(\ref{eq:SolvSp}). The colored domains correspond to the stable stationary solutions of Eq.~(\ref{eq:3krd}) with parameter values for which $\mathrm{Re}(\lambda)<0$.}
\label{fig2}
\end{figure}

 For $\alpha=0$, linear stability of the stationary solution $\mathbf{q_0}(\mathbf{r})$ can be analyzed by means of the ansatz $\mathbf{q}(\mathbf{r},\,t)=\mathbf{q_0}(\mathbf{r})+\boldsymbol{\varphi}(\mathbf{r})\exp{(\mu\,t)}$, leading to the linear eigenvalue problem
\begin{equation}\label{eq:EVPOD}
  \mathfrak{L}'(\mathbf{q_0})\boldsymbol{\varphi}=\mu\,\boldsymbol{\varphi}\,,
\end{equation}
where the linear operator $\mathfrak{L}'(\mathbf{q_0})$ stands for a linearization of the operator $\mathfrak{L}$ around the stationary solution $\mathbf{q_0}$, $\mu$ is the set of eigenvalues and $\boldsymbol{\varphi}(\mathbf{r})$ are the corresponding eigenfunctions. Notice that generally the operator $\mathfrak{L}'(\mathbf{q_0})$ is not self-adjoint, so its eigenvalues and eigenfunctions are typically complex. The destabilization scenario we are interested in here is that a pair of complex-conjugated eigenvalues passes through the imaginary axis as a control parameter exceeds some critical value. In what follows, we use the time constant $\theta$ as a control parameter. If $\theta$ is below a critical value $\theta_c$, both inhibitors almost instantaneously adapt to the current distribution of the activator and the stationary solution remains stable. Therefore a mechanism for imposing instabilities is delayed inhibition, where one or several inhibitors are too slow to adapt to activator changes~\cite{
PurwinsDS2010}. Indeed, if $\theta>\theta_c$, an addition of a symmetric perturbation in form of the unstable breathing mode to the localized structure leads to slightly increased concentration in the center, that is, both concentrations become narrower than their counterparts in the stationary solution. Therefore they spread to the sides due to diffusion. While the inhibitor is slow, it strongly increases its concentration to overcompensate the losses caused by diffusion, whereas the activator decreases only slightly. That is, after some time one gets excess amount of the inhibitor, distributed over the whole localized structure. It causes a decay of the whole solution, especially on the tails of the activator. Hence both peaks become narrower. On the other hand, the losses due to diffusion cause the decay of the inhibitor in the center.  As a consequence, the activator can again slightly increase its concentration, and new breathing cycles begins. Figure~\ref{fig1} shows the typical behavior of the 
localized structure beyond the Andronov-Hopf-bifurcation point $\theta=\theta_c$ calculated by a numerical integration of the system~(\ref{eq:3krd}). In Fig.~\ref{fig1}~(a), a stationary localized solution bifurcates to a breathing localized structure~\cite{GurevichPRE2006}, which oscillates with a constant amplitude. Figure~\ref{fig1}~(b) shows another possible scenario, where the amplitude of oscillations of the localized structure increases with time. Here, slow inhibitors cannot suppress a strong increase in the activator concentration in the course of time, leading to a collapse of the solution. The former case is referred to as supercritical Hopf bifurcation, whereas the latter scenario is called subcritical. 
We focus first on the supercritical case, that is, an increase of the control parameter beyond the critical value $\theta_c$ leads to formation of the breathing localized structure. Our goal now is to investigate the behavior of the breathing localized structure in the presence of the time delayed feedback term and explore whether this type of oscillating solution can be stabilized with this kind of control. The stationary localized solution $\mathbf{q_0}$ exists for all values of the delay strength $\alpha$ and is not affected by the delay term. However, its stability may change~\cite{GurevichPRE2013}. For $\alpha\neq 0$ the linear stability of $\mathbf{q_0}$ is given by the eigenvalue problem
\begin{equation}\label{eq:DelayedEV}
 \mathfrak{L}'(\mathbf{q_0})\,\boldsymbol{\varphi}=\biggl(\lambda-\alpha\,\bigl(1-e^{-\lambda\,\tau}\bigr)\biggr)\,\boldsymbol{\varphi}\,,
\end{equation}
with the same set of eigenfunctions $\boldsymbol{\varphi}$ as in (\ref{eq:EVPOD}), as the linearization operator commutes with the identity coupling matrix. The complex eigenvalues $\lambda$ can be found  in terms of the Lambert function $W_m$, $m\in\mathbb{Z}$~\cite{HoevelPRE2005, TlidiPRL2009, Gurevich13, Corless96} as
$$
\lambda=\mu+\alpha+\frac{1}{\tau}\,W_m\bigl(-\alpha\tau\,\exp\bigl(-(\mu+\alpha)\tau\bigr)\bigr)\,.
$$
A stabilization of the unstable localized solution can be achieved for such values of $\alpha$ and $\tau$, where $\mathrm{Re}(\lambda)$ is negative. Separating the real and the imaginary part of the last equation and solving the obtained system for $\mathrm{Re}(\lambda)=0$, the following solvability condition for the instability threshold can be derived:
\begin{equation}\label{eq:SolvSp}
 \pm \alpha\tau\,\sqrt{1-\left(1+\frac{\mathrm{Re}(\mu)}{\alpha}\right)^2}=\pm \mathrm{arccos}\left(1+\frac{\mathrm{Re}(\mu)}{\alpha}\right)-\mathrm{Im}(\mu)\,\tau+2\pi n\,,\quad n\in\mathbb{Z}.
\end{equation}
As $\mathrm{Re}(\mu)>0$ beyond the bifurcation point $\theta=\theta_c$, Eq.~(\ref{eq:SolvSp}) possesses nontrivial solutions only for negative values of the delay strength $\alpha$, whereas a maximal possible value of $\alpha$ to stabilize unstable periodic localized solution is given by $\alpha=-\mathrm{Re}(\mu)/2$. In order to find the shape of the domains of stabilization we solve the solvability condition~(\ref{eq:SolvSp}) numerically for different values of $\tau$ and $\alpha$ as shown in Fig.~\ref{fig2}.
The obtained bifurcation diagram clearly indicates the influence of the time delayed feedback term on the dynamics of the breathing localized solution: The colored domains appertain to $\mathrm{Re}(\lambda)\leq 0$, where the stabilization is successful, whereas outside of these domains no stabilization takes place and the induced instabilities are caused by eigenvalues with $\mathrm{Re}(\lambda)>0$ and (in general) with non-vanishing imaginary parts.  
\section{Direct numerical simulations}

In order to verify results obtained from linear stability analysis, direct numerical simulations of the time evolution of the localized breathing solution of the system~(\ref{eq:3krd}) have been performed. The system parameters were chosen to be in the supercritical regime (like in Fig.~\ref{fig1}~(a)). The calculations were performed on the rectangular domain $\Omega=[-L,\, L]\times[-L,\, L]$ with periodic boundary conditions using a pseudospectral method, whereas a Runge-Kutta 4 scheme is employed for the time
stepping.  At the simulation beginning, the feedback term was switched off, that is, a breathing localized structure, located in the center of the domain $\Omega$ and widely separated from its boundaries, emerges. As soon as the amplitude of oscillations reaches the constant value, the time delayed term was switched on. Notice that as the localized solution is situated far from domain boundaries, boundary interaction effects are infinitesimal and play no role in the dynamical behavior. In the case of Neumann boundary conditions boundary effects are negligible, too, for the same geometrical settings.
\begin{figure}[!h]
\centering
 \includegraphics[width=0.95\textwidth]{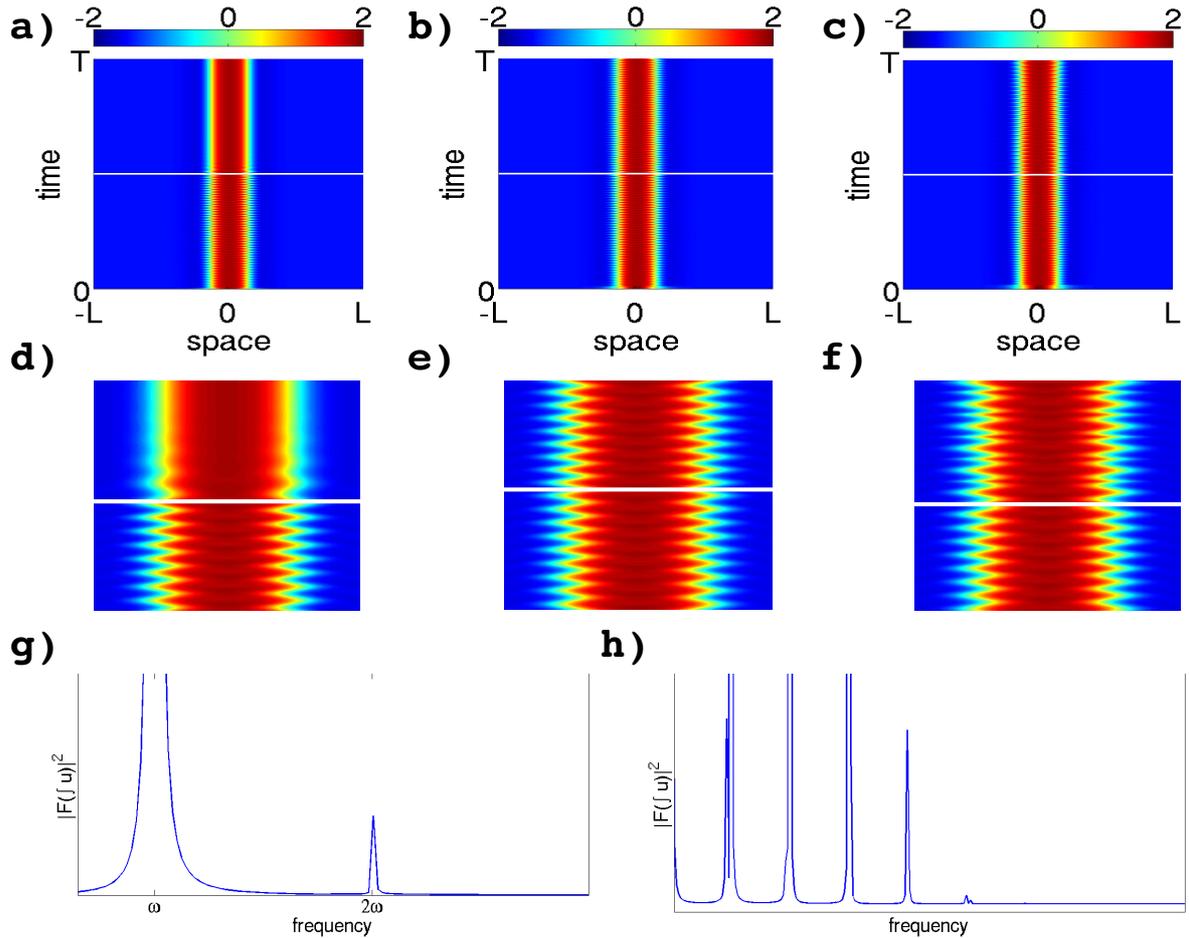} 
\caption{Time-delayed feedback control of the breathing localized structure in a supercritical regime. (a--f): Space-time representation of the activator distribution $u$ for different values of $\alpha$ and $\tau$ obtained from numerical solution of Eqn.~(\ref{eq:3krd}) is presented. White line indicates a time moment, where the time delayed feedback control is applied. (a) $\alpha=-0.3$, $\tau=2$: Stabilization is successful; (b) $\alpha=-0.3$, $\tau=8$: The control leads to a breathing periodic solution; (c) $\alpha=-3$, $\theta=14$: The control leads to a breathing quasiperiodic solution. The panel (d-f) shows corresponding zoom views of the region, where the control is applied. The bottom panel (g,\,h) shows the power spectra calculated for induced breathing (b) as well as for the delay induced quasiperiodic breathing solution (c). Other parameters: $D_u=4.7\cdot10^{-3}$, $D_v=0$, $D_w=0.01$, $\lambda=5.67$, $\kappa_1=-1.04$, $\kappa_3=1.0$, $\kappa_4=3.33$, $\eta=0.6$, $\theta=0.58$. The calculations 
were performed on the rectangular domain $\Omega=[-L,\, L]\times[-L,\, L]$, $L=1.0$ with periodic boundary conditions. }
\label{fig3}
\end{figure}
Three simulation examples are presented in Fig.~\ref{fig3}, where space-time plots for different values of $\alpha$ and $\tau$ are shown (top panel (a-c)). A white line indicates the time moment where the time-delayed term was switched on. The panel (d-f) represents zoom views of the regions, where delayed feedback is applied. One can see that for parameters within stabilization domains, the amplitude of oscillations decreases after the activation of the time delayed term and stationary localized structure emerges (Fig.~\ref{fig3}~(a,\,d)). On the other hand, on the outside of these domains, no stabilization takes place and solutions, oscillating with slightly different amplitude can be found (see Fig.~\ref{fig3}~(b,\,e)). However, the induced oscillatory instability scenario can be more complex as shown in Fig.~\ref{fig3}~(c,\,f), where a quasiperiodic oscillating localized structure arises as time delay term is switched on. In order to characterize the behavior of the obtained breathing and quasiperiodic 
breathing solutions, a power spectrum $|F(\int_{\Omega} u(\mathbf{r},\,t))|^2$ in terms of Fourier transform $F$ was calculated for the activator component $u(\mathbf{r},\,t)$ (see Fig.~\ref{fig3}~(bottom panel (g,\,h))). Here, one can clearly see the difference between power spectra of a single breathing solution presented in Fig.~\ref{fig3}~(b,\,e) and a quasi-periodic breathing (Fig.~\ref{fig3}~(c,\,f)). In the first case, the periodic signal gives peaks at a fundamental frequency $\omega$ and its harmonics as shown in Fig.~\ref{fig3}~(g)). In the case of quasiperiodic breathing, several peaks at linear combinations of two or more irrationally related frequencies can be observed (see Fig.~\ref{fig3}~(h)).

\section{Order parameter equation}
Our next goal is to try to understand the results obtained from the linear stability analysis and supported by numerical simulations of the system in question from the point of view of bifurcation theory. As we mentioned above, the instability scenario we are interested in corresponds to the situation, where a pair of complex-conjugated eigenvalues passes through the imaginary axis as one gradually changes the control parameter $\theta$. That is, for some critical value $\theta=\theta_c$, the corresponding eigenvalues are purely imaginary, i.e., $\lambda=\pm i\omega$. Now, if we increase the control parameter $\theta=\theta_c+\varepsilon$, $\varepsilon\ll 1$ the real-valued vector function $\mathbf{q}(\mathbf{r},\,t)$ can be represented as~\cite{GurevichPRE2006,GuFrMMNP13}
\begin{equation}\label{eq:ansatz}
 \mathbf{q}(\mathbf{r},\,t)=\mathbf{q_0}(\mathbf{r})+\boldsymbol{\xi}(t)\,\boldsymbol{\varphi}(\mathbf{r})\,e^{i\omega t}+\boldsymbol{\xi}_2(\mathbf{r},\,t)\,e^{2i\omega t}+\boldsymbol{\xi}_0(\mathbf{r},\,t)+\mathrm{c.c.}\,.
\end{equation}
Here,  $\boldsymbol{\xi}(t)$ is a slow varying complex amplitude of the critical eigenfunction $\varphi(\mathbf{r})$, corresponding to the eigenvalue $\lambda=\pm i\omega$ at the bifurcation point $\theta= \theta_c$. In addition, $\boldsymbol{\xi}_2(\mathbf{r},\,t)$ and $\boldsymbol{\xi}_0(\mathbf{r},\,t)$ stay for the contribution of the second and zero harmonics, respectively. The specific choice of the perturbation $\widetilde{\mathbf{q}}(\mathbf{r},\,t)=\mathbf{q}(\mathbf{r},\,t)-\mathbf{q_0}(\mathbf{r})$ in the ansatz~(\ref{eq:ansatz}) becomes apparent if one takes a look at the power spectrum $|F(\int_{\Omega} \widetilde{\mathbf{q}}(\mathbf{r},\,t))|^2$ calculated by means of Fourier transform $F$ (see Fig.~\ref{fig1}~(c), where power spectrum for the activator perturbation $u-u_0$ is shown). Here, one can clearly see that apart of the fundamental frequency $\omega$, only zero and second harmonics impact on the spectrum of the oscillating solution.
Our goal is to write down an ordinary differential equation for the amplitude $\boldsymbol{\xi}(t)$ of the unstable mode $\boldsymbol{\varphi}(\mathbf{x})$, which describes the behavior of the single localized structure in the vicinity of the Andronov-Hopf-bifurcation point. For this purpose we substitute Eq.~(\ref{eq:ansatz}) into Eq.~(\ref{eq:GenEqDel}), equalize the terms with the frequency $\omega$ and obtain
\begin{equation}
 \dot{\boldsymbol{\xi}}\,\boldsymbol{\varphi}=\varepsilon\,\boldsymbol{\xi}\mathcal{L}'_{\varepsilon}\boldsymbol{\varphi}+\boldsymbol{\xi}\,\mathcal{L}''_c\boldsymbol{\varphi}\,\boldsymbol{\xi}_0+\overline{\boldsymbol{\xi}}\,\mathcal{L}''_c\overline{\boldsymbol{\varphi}}\,\boldsymbol{\xi}_2+\frac{1}{2}\,\mathcal{L}'''_c\boldsymbol{\varphi\varphi}\overline{\boldsymbol{\varphi}}\,|\boldsymbol{\xi}|^2\,\boldsymbol{\xi}+\alpha\,\left(\boldsymbol{\xi}(t)-\boldsymbol{\xi}(t-\tau)\,e^{-i\omega\tau}\right)\,.
\end{equation}
Here, $\mathcal{L}'_{\varepsilon}=\frac{\partial\mathfrak{L}'(\mathbf{q}_0,\nabla,\theta)}{\partial\theta}\bigl|_{\theta=\theta_{c}}$ and $\mathcal{L}^{(n)}_c=\mathfrak{L}^{(n)}(\mathbf{q}_0,\nabla,\theta_{c})$, so that
$$
\mathfrak{L}'(\mathbf{q}_0,\nabla,\theta_{c}+\varepsilon)=\mathcal{L}'_c+\varepsilon\mathcal{L}'_{\varepsilon}\,,
$$
whereas the overline stands for complex conjugate. In addition, equalizing the terms with the frequencies $2\omega$ and zero, respectively, and neglecting contributions of the delayed terms, one gets two equations describing the time evolution of the amplitudes $\boldsymbol{\xi}_2$ and $\boldsymbol{\xi}_0$ of the stable modes as
\begin{eqnarray}
 \dot{\boldsymbol{\xi}}_2+2i\omega\boldsymbol{\xi}_2&=&\mathcal{L}'_c\boldsymbol{\xi}_2+\frac{\boldsymbol{\xi}^2}{2}\,\mathcal{L}''_c\boldsymbol{\varphi\varphi}\,, \label{eq:stabspec1}\\
  \dot{\boldsymbol{\xi}}_0&=&\mathcal{L}'_c\boldsymbol{\xi}_0+|\boldsymbol{\xi}|^2\,\mathcal{L}''_c\boldsymbol{\varphi}\overline{\boldsymbol{\varphi}}\,. \label{eq:stabspec2}
\end{eqnarray}
As we suppose that all critical modes except for the breathing mode $\boldsymbol{\varphi}$ are stable, we can apply the adiabatic approximation to Eqs.~(\ref{eq:stabspec1}-\ref{eq:stabspec2}), yielding two solvability conditions with respect to unknown functions $\boldsymbol{X}_2$ and $\boldsymbol{X}_0$ 
\begin{eqnarray}
 \left(2i\omega-\mathcal{L}'_c\right)\boldsymbol{X}_2&=&\mathcal{L}''_c\boldsymbol{\varphi\varphi}\,,\\
 \mathcal{L}'_c\boldsymbol{X}_0&=&-\mathcal{L}''_c\boldsymbol{\varphi}\overline{\boldsymbol{\varphi}}\,.
\end{eqnarray}
Notice that Eqs.~(\ref{eq:stabspec1}-\ref{eq:stabspec2}) were obtained neglecting contributions of the delayed terms. However, this simple approximation may be reasonable, while adiabatic elimination excludes from consideration the time evolution of stable modes. The functions $\boldsymbol{X}_2$ and $\boldsymbol{X}_0$ are living in the same space as $\boldsymbol{\varphi}$ and are connected to the amplitudes $\boldsymbol{\xi}_2$ and $\boldsymbol{\xi}_0$ as~\cite{Friedrich2005, Haken}
$$
\boldsymbol{\xi}_2=\boldsymbol{X}_2\boldsymbol{\xi}^2\,\quad \boldsymbol{\xi}_0=\boldsymbol{X}_0\,|\boldsymbol{\xi}|^2\,.
$$
Now, in order to write the desired equation for $\boldsymbol{\xi}$, we need to project onto the corresponding mode $\boldsymbol{\varphi}^{\dag}$ of the adjoint operator $\mathcal{L}'^{\dag}_c$. The projection yields:

\begin{equation}\label{eq:OPE}
 \partial_{t}\boldsymbol{\xi}=\varepsilon\,a_1\,\boldsymbol{\xi}+a_2\,\boldsymbol{\xi}|\boldsymbol{\xi}|^2+\alpha\,\left(\boldsymbol{\xi}(t)-\boldsymbol{\xi}(t-\tau)\,e^{-i\omega\tau}\right)\,,
\end{equation}
where complex coefficients $a_1$ and $a_2$ can be expressed as
$$
a_1=\frac{\langle\boldsymbol{\varphi}^{\dag}|\mathcal{L}'_{\varepsilon}\boldsymbol{\varphi}\rangle}{\langle\boldsymbol{\varphi}^{\dag}|\boldsymbol{\varphi}\rangle}\,,\quad a_2=\frac{\langle\boldsymbol{\varphi}^{\dag}|\mathcal{L}'''_c\boldsymbol{\varphi\varphi}\overline{\boldsymbol{\varphi}}\rangle}{2\,\langle\boldsymbol{\varphi}^{\dag}|\boldsymbol{\varphi}\rangle}+\frac{\langle\boldsymbol{\varphi}^{\dag}|\mathcal{L}''_c\mathbf{X}_0\boldsymbol{\varphi}\rangle+\langle\boldsymbol{\varphi}^{\dag}|\mathcal{L}''_c\boldsymbol{X}_2\
\overline{\boldsymbol{\varphi}}\rangle}{\langle\boldsymbol{\varphi}^{\dag}|\boldsymbol{\varphi}\rangle}\,.
$$
Here, $\langle\cdot | \cdot\rangle$ denotes the scalar product defined in terms of full spatial integration over the considered domain.

Note that in general the analytical calculation of the eigenfunctions $\boldsymbol{\varphi}^{\dag}$ of the adjoint operator $\mathfrak{L}'^{\dag}(\mathbf{q_0})$ is difficult, but in the case of the reaction-diffusion system~(\ref{eq:3krd}) it is possible using the relation~\cite{MoskalenkoEPL2003,Gurevich2004}
\begin{equation}\label{eq:condM}
\boldsymbol{\varphi}^{\dag}=M(\eta,\,\theta)^{-1}\,\overline{\boldsymbol{\varphi}}\,, 
\end{equation} 
where $M(\eta,\,\theta)$ is a diagonal matrix, defined as
$$
M(\eta,\,\theta)=\begin{pmatrix}
       1 & 0 & 0\\
       0 & -\frac{1}{\kappa_3\,\eta} & 0\\
       0 & 0 & -\frac{1}{\kappa_4\theta}
  \end{pmatrix}\,.
$$
For example, the coefficient, standing in the numerator of $a_1$, can be calculated as~\cite{GurevichPRE2006}
$$
\langle\boldsymbol{\varphi}^{\dag}|\mathcal{L}'_{\varepsilon}\boldsymbol{\varphi}\rangle=i\omega\,\langle M_c^{-1}\overline{\boldsymbol{\varphi}}|M_{\theta}M_{c}^{-1}\boldsymbol{\varphi}\rangle=i\omega\kappa_4\,\langle\varphi_w^2\rangle\,,
$$
where $M_c=M(\eta,\,\theta_c)$ and $M_{\theta}=\frac{\partial M(\eta,\,\theta)}{\partial\theta}\bigl|_{\theta=\theta_c}$. That is, using the relation~\eqref{eq:condM} one can calculate all scalar products in Eq.~\eqref{eq:OPE} in terms of the critical eigenfunction $\boldsymbol{\varphi}=\left(\varphi_u,\,\varphi_v,\,\varphi_w\right)^T$ of the linearization operator $\mathcal{L}'_c$.

A nonlinear delay-differential equation~(\ref{eq:OPE}) is the desired order parameter equation, describing the behavior of a single localized structure in the vicinity of the bifurcation point $\theta=\theta_c$. Notice that a similar equation was obtained in~\cite{Landa} for a model of Doppler's autodyne, described by the van der Pol-Duffing generator with additional delayed feedback.  

For the vanishing feedback force ($\alpha=0$), Eq.~(\ref{eq:OPE}) is reduced to a normal form of a Hopf bifurcation~\cite{Kuramoto,GurevichPRE2006}. That is, for $\alpha=0$ a trivial solution $\boldsymbol{\xi}=0$, corresponding to the case of a stable stationary localized structure, is stable for $\mathrm{Re}(a_1)<0$, whereas for $\mathrm{Re}(a_1)>0$ a nontrivial periodic solution $\boldsymbol{\xi}(t)=R_0\,e^{i\nu_0t}$ can be found,
$$
R_0=\sqrt{-\varepsilon\frac{\mathrm{Re}(a_1)}{\mathrm{Re}(a_2)}}\,,\quad \nu_0=\varepsilon\,\mathrm{Im}(a_1)+R_0^2\,\mathrm{Im}(a_2)
$$
which is stable for $\mathrm{Re}(a_2)<0$ and unstable for $\mathrm{Re}(a_2)>0$~\cite{Kuramoto,GurevichPRE2006}. The former case corresponds to supercritical Hopf bifurcation, where the growth of the unstable mode is stabilized by a limit cycle as shown in Fig.~\ref{fig1}~(a), whereas in the latter case the bifurcation is subcritical (see Fig.~\ref{fig1}~(b)). 

\begin{figure}[!h]
\begin{tabular}{ll}
(a) & (b)\\
 \includegraphics[width=0.5\textwidth]{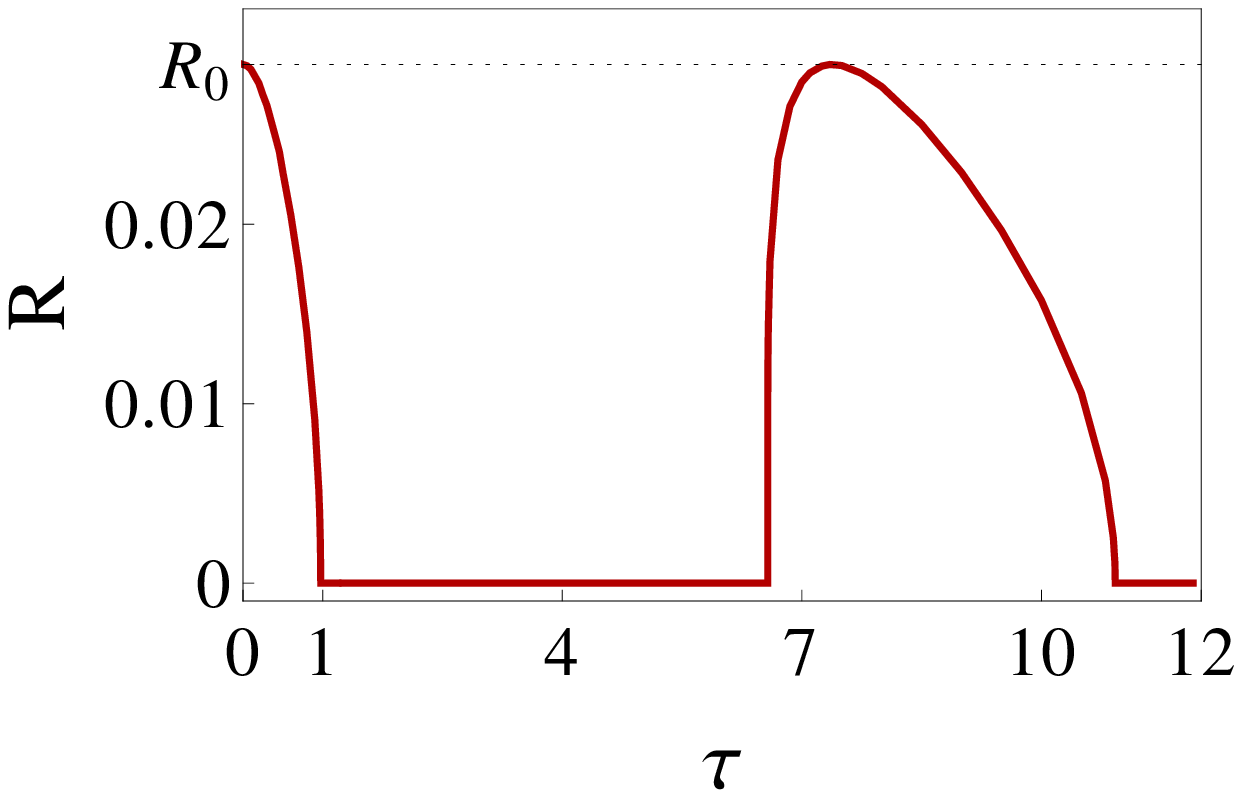} &\includegraphics[width=0.43\textwidth]{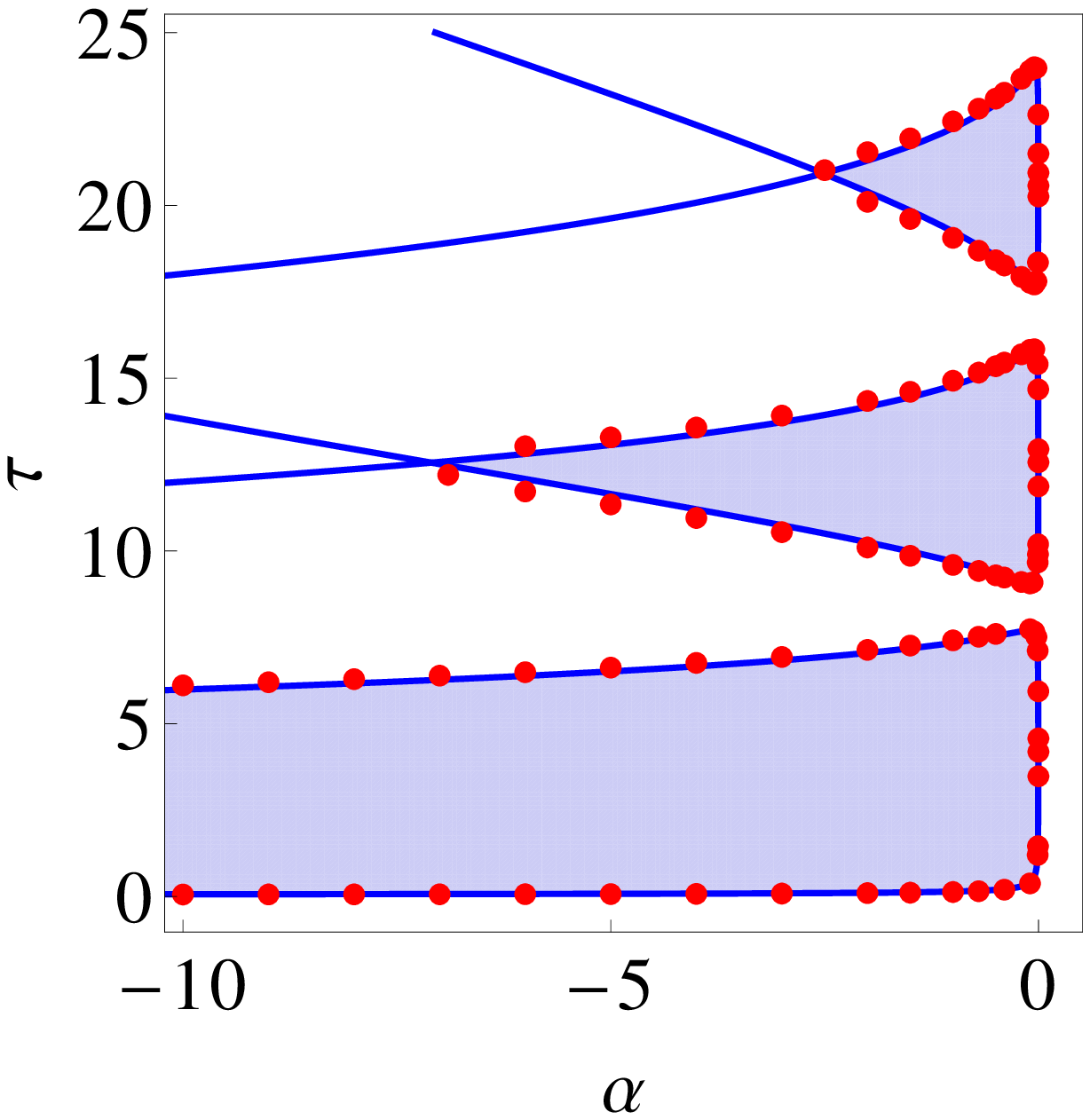}
\end{tabular}
\caption{(a) Dependence of the radius of the limit cycle $R=R_c$ on the delay time $\tau$ obtained from Eqs.~(\ref{eq:OPE21}-\ref{eq:OPE22}) for fixed value of the delay strength $\alpha=-0.2$; (b) The critical delay time given by Eq.~(\ref{eq:CritTau}) in dependence on delay strength $\alpha$ (blue solid lines). Red points correspond to stabilization threshold obtained from the solvability condition~(\ref{eq:SolvSp}), whereas colored regions indicate areas in $(\alpha\,,\tau)$ plane, where stabilization of the breathing localized structure is possible.}
\label{fig4}
\end{figure}

As indicated above, we are interested in the supercritical case, that is, in what follows $\mathrm{Re}(a_1)>0$ and $\mathrm{Re}(a_2)<0$. Now our aim is to find out the influence of the time delayed feedback term on the dynamics of the periodic solution, given by the latter relation. For non-vanishing feedback strength, $\alpha\neq 0$, the exponential ansatz for the complex amplitude, $\boldsymbol{\xi}(t)=R\,e^{i\phi}$, leads to the following system of coupled nonlinear real-valued delay differential equations
\begin{eqnarray}\label{eq:OPE11}
    \dot{R}&=&\varepsilon\,\mathrm{Re}(a_1)\,R+\mathrm{Re}(a_2)\,R^3+\alpha\left(R-R(t-\tau)\cos(\phi-\phi(t-\tau))\right)\,,\\
     \dot{\phi}R&=&\varepsilon\,\mathrm{Im}(a_1)\,R+\mathrm{Im}(a_2)\,R^3+\alpha\,R(t-\tau)\sin(\phi-\phi(t-\tau))\,. \label{eq:OPE12}
  \end{eqnarray}
 This system possesses a stationary solution
  $$
  R=R_c,\quad \phi=\vartheta\,t+\vartheta_0\,,
  $$
  corresponding to a periodic motion on a limit cycle of the radius $R=R_c$. Here, $\vartheta_0$ is an arbitrary constant and 
\begin{eqnarray}\label{eq:OPE21}
    R_c^2&=&\frac{-\varepsilon\,\mathrm{Re}(a_1)}{\mathrm{Re}(a_2)}+\frac{\alpha\left(\cos((\vartheta+\omega)\,\tau)-1\right)}{\mathrm{Re}(a_2)}\,,\\
    \vartheta&=&\varepsilon\,\mathrm{Im}(a_1)+\mathrm{Im}(a_2)\,R_c^2+\alpha\,\sin((\vartheta+\omega)\,\tau)\,. \label{eq:OPE22}
   \end{eqnarray}
  One can easily show that for any fixed value of the delay strength $\alpha$, the maximal value of the radius $R_c$ as a function of the delay time $\tau$ is achieved at $\tau=\pi\,k/(\vartheta+\omega)$, $k\in\mathbb{Z}$, what corresponds to the radius $R_0$ for vanishing delay term. That is, for these values of $\tau$, the time delayed feedback control procedure fails. However, for other values of $\tau$ and $\alpha$, Eqs.~(\ref{eq:OPE21}-\ref{eq:OPE22}) allow the control of the value of $R_c$ in a straightforward manner: Indeed, the value of $R_c$ is smaller then $R_0$ and can be directly calculated from system~(\ref{eq:OPE21}-\ref{eq:OPE22}), see Fig.~\ref{fig4}~(a), where the dependence of the radius $R=R_c$ on the delay time $\tau$ is shown for $\alpha=-0.2$.
 
Here, nontrivial values of $R_c$ correspond to a localized structure, breathing with a constant amplitude (compare with Fig.~\ref{fig3}~(b)), whereas the vanishing $R_c$ correspond to the stabilization of the breathing due to time-delayed feedback (see Fig.~\ref{fig3}~(a)). Solving system ~(\ref{eq:OPE21}-\ref{eq:OPE22}) for $R_c=0$, the critical delay time $\tau$, which is necessary to archive stabilization, can be explicitly found as 
\begin{equation}\label{eq:CritTau}
 \tau=\frac{\pm\mathrm{arccos}\left(1+\frac{\varepsilon\mathrm{Re}(a_1)}{\alpha}\right)+2\pi n}{\varepsilon\,\mathrm{Im}(a_1)+\omega\pm\alpha\,\sqrt{1-\left(1+\frac{\mathrm{Re}(a_1)}{\alpha}\right)^2}}\,.
\end{equation}
One can see that for a given value of $\alpha$, the critical delay time depends on the parameters $a_1$ and $\omega$, defined at the bifurcation point $\theta_c$ as well as on the distance to the bifurcation point $\varepsilon$. That is, the coefficients of the order parameter equation~(\ref{eq:OPE}), derived at the bifurcation point, provide the full information concerning stabilization threshold~(\ref{eq:CritTau}) in the vicinity of the bifurcation point without any need to calculate the spectrum of the linearization operator like we did for the solvability condition~(\ref{eq:SolvSp}). 

Figure~\ref{fig4}~(b) shows the critical delay time~(\ref{eq:CritTau}) in dependence on the delay strength $\alpha$ (blue solid lines), whereas red filled dots correspond to stabilization threshold obtained from the solvability condition~(\ref{eq:SolvSp}). Here, colored regions indicate areas in $(\alpha\,,\tau)$ plane, where stabilization of the breathing localized structure is possible, i.e., the amplitude of the oscillations $R$ is zero inside of colored domains and equals $R_c$ outside. However, outside of the stabilization domains the stationary solution~(\ref{eq:OPE21}-\ref{eq:OPE22}) of system~(\ref{eq:OPE21}-\ref{eq:OPE22}) can become unstable as one  changes delayed feedback parameters. This situation corresponds to the instability scenario shown in Fig.~\ref{fig3}~(c), where the switching on of the control force leads to a complex quasiperiodic oscillations of the localized structure. 

Heretofore, the control and stabilization of the supercritical regime ($\mathrm{Re}(a_1)>0$, $\mathrm{Re}(a_2)<0$ for $\alpha=0$) was discussed. However, the stationary solution~(\ref{eq:OPE21}-\ref{eq:OPE22}) can also be found for a proper choice of $\alpha$ and $\tau$ even if $\mathrm{Re}(a_2)>0$ for $\alpha=0$. 
\begin{figure}[!h]
\centering
\begin{tabular}{ll}
(a) & (b)\\
 \includegraphics[width=0.38\textwidth]{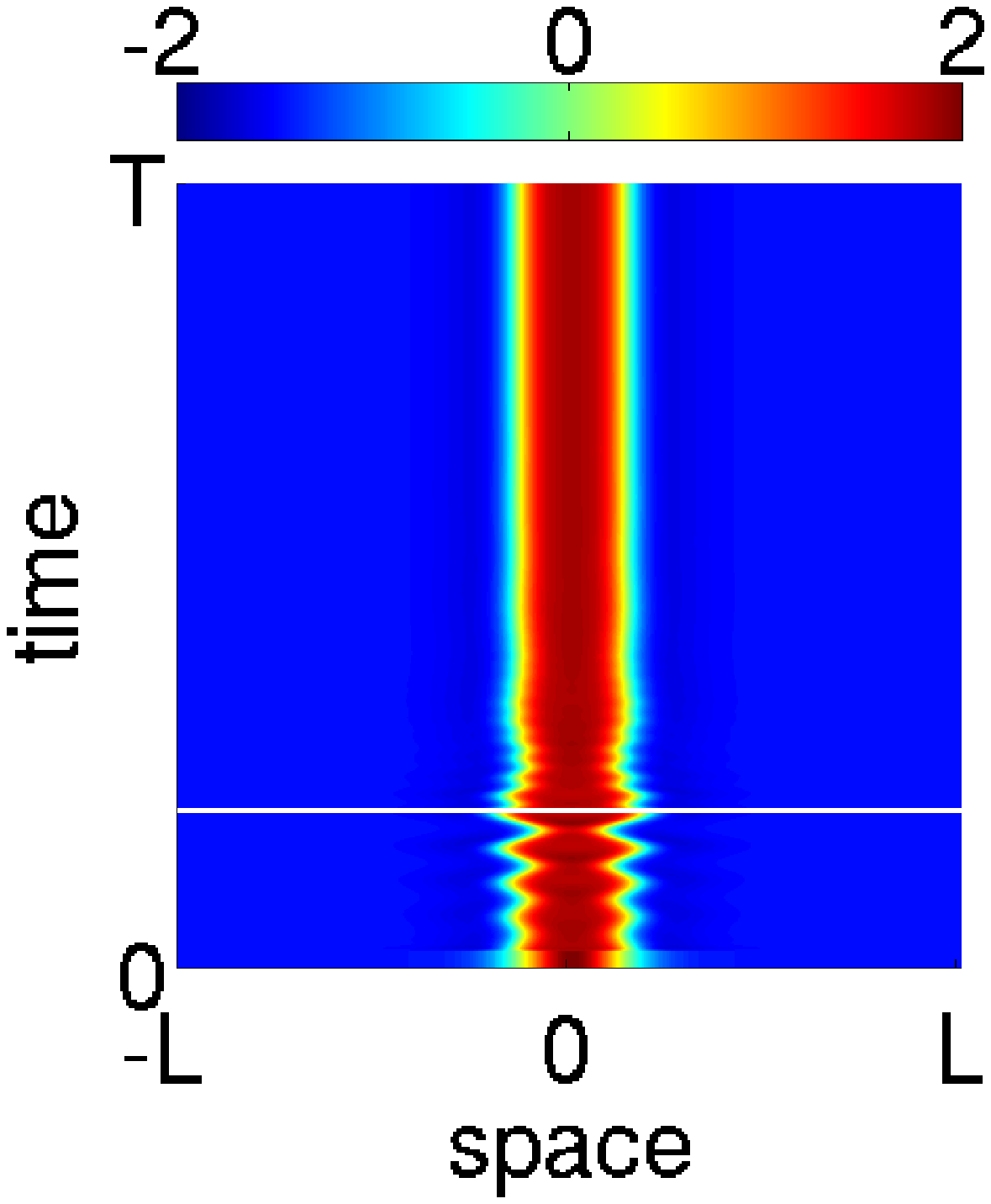} &\includegraphics[width=0.37\textwidth]{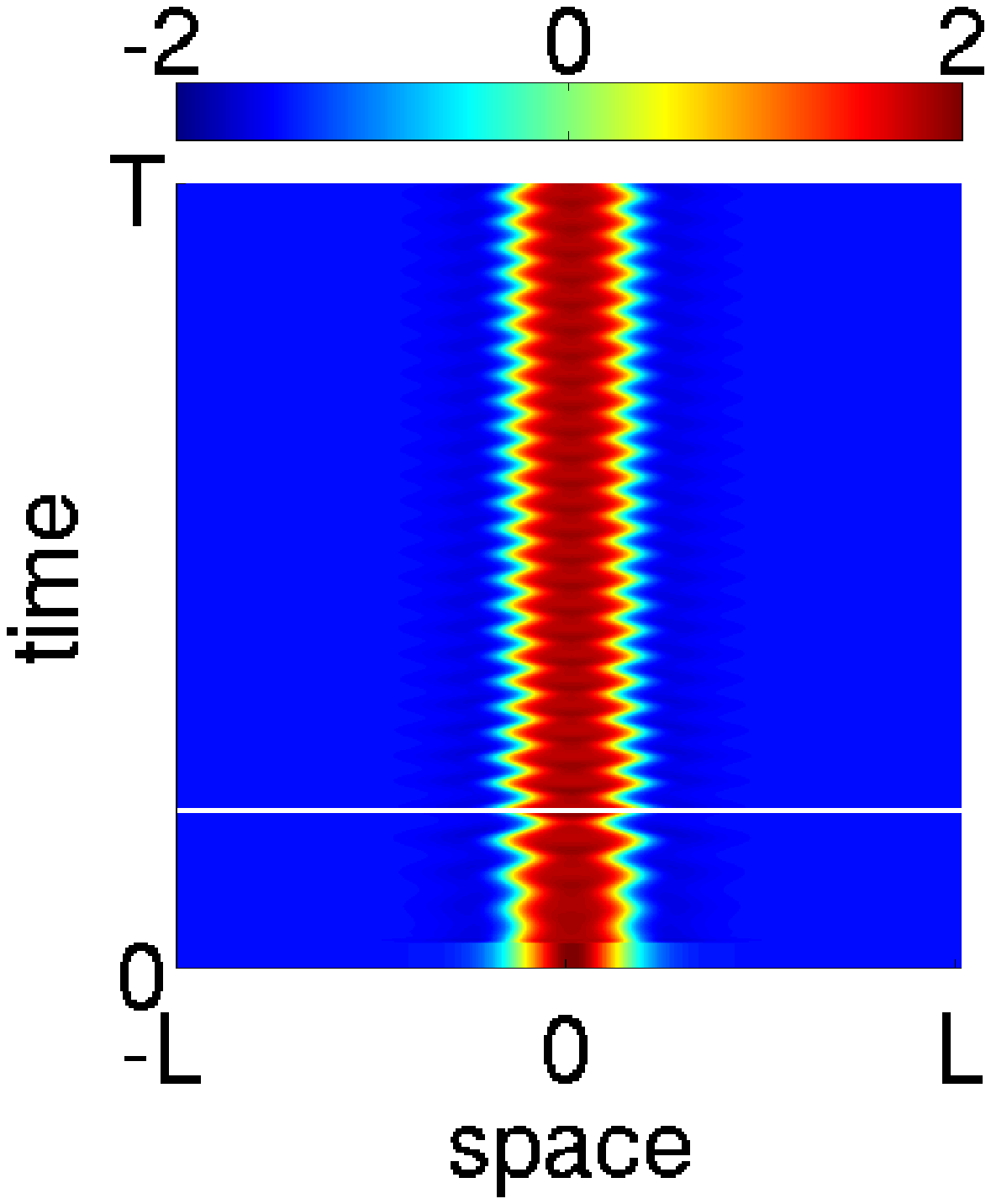}
\end{tabular}
\caption{Control of the breathing localized structure in a subcritical regime. Space-time representation of the activator distribution $u$ for different values of $\alpha$ and $\tau$ obtained from numerical solution of Eqn.~(\ref{eq:3krd}) is shown. White line indicates a time moment, where the time delayed feedback control is switched on. (a) $\alpha=-0.5$, $\tau=4$: The control leads to stabilization of the solution ; (b) $\alpha=-0.5$, $\tau=6$: The control leads to the formation of breathing localized structure. Other parameters: $D_u=4.7\cdot10^{-3}$, $D_v=0$, $D_w=0.01$, $\lambda=5.67$, $\kappa_1=-1.04$, $\kappa_3=1.0$, $\kappa_4=3.33$, $\eta=0.62$, $\theta=0.64$. The calculations were performed on the rectangular domain $\Omega=[-L,\, L]\times[-L,\, L]$, $L=1.0$ with periodic boundary conditions. 
}
\label{fig5}
\end{figure}
This situation is presented in Fig.~\ref{fig5}, where a direct numerical simulation of Eqn.~(\ref{eq:3krd}) is performed for parameters in the subcritial regime (see also Fig.~\ref{fig1}~(b)). Here, a single localized structure  in the absence of nonlinear stabilization is prevented from a collapse, that leads either to a stabilization of the oscillations (Fig.~\ref{fig5}~(a)) or to a breathing with a constant amplitude (Fig.~\ref{fig5}~(b)). As was mentioned above, the collapse of the localized structure in the absence of control is caused by the delayed inhibition, that is, both inhibitors are too slow to suppress a strong increase in the activator concentration in the course of time. In this framework, the time delayed feedback control can be seen as an additional delayed inhibition, forcing slow inhibitors to be fast enough to follow the activator distribution. Using a concept of additional induced delayed inhibition, all results, obtained for the supercritical case can be explained in the same way: The 
time delayed feedback control provides a mechanism of additional delayed inhibition, that can either lead to the stabilization of the solution if controlled inhibitors are fast enough, or to oscillatory dynamics otherwise.

\vspace*{-10pt}

\section{Conclusion}

In the present paper we have discussed the dynamics of breathing localized structures in a three-component reaction-diffusion system under time delayed feedback control. We have analytically studied the linear stability problem of the delayed system yielding explicit expression for the boundary of stabilization domains, within a breathing localized structure can be effectively stabilized. However, we have found that outside of these domains more complex delay-induced periodic or quasiperiodic oscillations can be obtained. In order to understand the influence of the delayed feedback term on the behavior of the breathing localized structure, an order parameter equation for the amplitude of the breathing mode was derived. The information about dynamics of the system is contained in the complex coefficients of this equation. Using this information, the dependence of the oscillation radius on delay parameters can be explicitly derived, providing a robust mechanism to control the behavior of the breathing 
localized structure in a straightforward manner.
Notice that obtained order parameter equation is different from the normal form of the (subcritical) Hopf bifurcation subjected to the time-delayed feedback, which contains a complex phase ahead of the delayed term~\cite{FiedlerPRL2007, JustPRE2007}. However, it falls into place if one recollects that the reaction-diffusion system~(\ref{eq:3krd}) is a real-valued system with real-valued delay rate. That is, one may obtain the normal form, mentioned above, considering complex-valued model systems (e.g., a complex Ginzburg-Landau equation) with complex delay rates. To conclude, note that although all results are obtained for the identity coupling term, the form of order-parameter equation still stand for other coupling matrices. However, explicit calculation of the corresponding coefficients in this case is intricate and will be treated elsewhere. Moreover, all results are derived in general form and can be applied to a wide class of spatial extended non-variational systems admitting localized structures like 
control of breathing and spatiotemporal chaotic localized states discussed in the framework of the non-variational Swift-Hohenberg equation and also observed in liquid crystal light value experiments ~\cite{VeschuerenPRL2013,ClercPRE2013}.


 \end{document}